\documentclass
[preprint,a4paper,byrevtex,nofootinbib,nobibnotes,noshowpacs,prb]{revtex4}%
\usepackage{amsfonts}
\usepackage{amssymb}
\usepackage{amsmath}
\usepackage{graphicx}%
\begin{document}
\author{Xin Zhou*\footnotetext{* Corresponding author. E-mail: xinzhou@wipm.ac.cn;
Tel: +86-27-8719-7796; Fax: +86-27-8719-9291.}, Jun Luo, Xian-ping Sun, Xi-zhi
Zeng, Shang-wu Ding, Mai-li Liu, Ming-sheng Zhan}
\affiliation{State Key Laboratory of Magnetic Resonance and Atomic and Molecular
Physics,\ Wuhan Institute of Physics and Mathematics, The Chinese Academy of
Sciences, P. O. Box 71010, Wuhan 430071, People's Republic of China}
\title{Enhancement of solid-state proton NMR via SPINOE with laser-polarized xenon}

\begin{abstract}
We have first successfully transferred the $^{129}$Xe polarization of natural
isotopic composition to the proton of solid-state $^{1}$HCl via Spin
Polarization-Induced Nuclear Overhauser Effect (SPINOE), by mixing the
hyperpolarized $^{129}$Xe gas and the $^{1}$HCl gas and then cooling them to
their condensated state in a flow system. The solid-state enhanced factor of
the NMR signal of 6 for $^{1}$H was observed, and the equation of solid-state
polarization enhancement via cross relaxation has also been theoretically
deduced. Using this equation, the theoretically calculated enhancement is in
agreement with the measured value within error. Also, this technique is maybe
useful to establish a solid state NMR quantum computer.

PACS number(s): 32.80.Bx, 33.25.+k, 03.67.Lx

\end{abstract}
\maketitle

\bigskip\bigskip\newpage

\section{Introduction}

\ \ \ Many difficulties appear in scaling the liquid-state Nuclear Magnetic
Resonance (NMR) quantum computers to multi-qubit, since the signal intensity
is low sensitive and decreases exponentially with increasing the number of
qubits. Therefore, the maximal number of qubits of quantum computer using the
current liquid-state NMR technology might be limited to 10 \cite{Warren}.
Although there are some methods developed for enhancing polarization, such as
lower temperature and higher magnetic fields, Dynamic Nuclear Polarization
(DNP) \cite{Abragam,Bates}, Nuclear Overhauser Effect (NOE) \cite{Overhauser},
cross polarization \cite{Slichter}, etc., unfortunately these approaches can
provide only limited relief or need rigid conditions for polarization
enhancement. A promising route to enhance the nuclear polarization is the
utilization of optical pumping and spin exchange \cite{Happer84,Zeng,Happer},
which can increase the nuclear polarization by four or five orders of
magnitude over thermal equilibrium. The hyperpolarized $^{129}$Xe and $^{3}$He
gases are greatly effective in the development of magnetic resonance imaging
\cite{Albert}, surface science \cite{Pietrass}, probing of biological system
\cite{Rubin,Cherubini}, polarized targets \cite{Xu}, neutron polarization
\cite{Jones}, precision measurement \cite{Bear}, polymer
science\cite{Nagasaka}, quantum computation \cite{Chuang}, etc..

Hitherto, proton sensitivity enhancements are not very large in all
experiments \cite{Song,Navon,Fitzgerald,Room}\ except for surface enhancements
\cite{Happer93}, and all of these experiments have almost been implemented by
using gaseous or liquid hyperpolarized xenon. The $^{1}$H polarization has
been enhanced by a factor of 0.1 to 2 on a 4.2T NMR spectrometer at room
temperature via cross-relaxation between dissolved hyperpolarized gaseous
$^{129}$Xe and $^{1}$H of the liquid benzene solvent, which has also been
firstly called the SPINOE by Pines' group\cite{Navon}. Through dissolving
compounds in hyperpolarized liquid xenon, the enhanced signal of over 45 for
$^{1}$H has also been observed at 1.4T and 200K by Happer's group
\cite{Fitzgerald}. In Xe ice, larger nuclear polarization of hyperpolarized
$^{129}$Xe has been transferred to $^{13}$CO$_{2}$ by low-field thermal mixing
\cite{Bowers}. However, as far as we know, solid-state proton polarization
enhancement has not been reported in the literature.

Solid-state quantum computers, which could execute the 10$^{6}$ qubits
operation brought out by DiVincenzo \textit{et al}., were widely noticed
\cite{DiVincenzo}. Considering that hydrogen chloride (HCl) and xenon have
almost the same melting points between 160 and 170K, and the longitudinal
relaxation time (T$_{1}$) of hyperpolarized $^{129}$Xe in solid-state (4K) is
about 500 hours \cite{Gatzke}, we put forward the idea of laser-enhanced
solid-state quantum computer (LESSQC) by mixing hyperpolarized $^{129}$Xe and
$^{1}$HCl.

In this letter, we firstly demonstrate that the polarization of hyperpolarized
solid-state $^{129}$Xe produced by spin exchange is transferred to the proton
of solid-state $^{1}$HCl via SPINOE without using low-field thermal mixing or
Hartmann-Hahn matching condition. This method yields the proton enhancement of
6 by comparison of that without optical pumping on a Bruker SY-80M NMR
spectrometer (1.87T) at 142 K, leading us to take a step towards realizing LESSQC.

\section{Experiment and Results}

The experimental apparatus, similar to that in Ref. 26, is shown schematically
in Figure 1. Briefly, the whole system consists of an optical pumping system
(the right part of the valve K4), which is used to produce the hyperpolarzied
gaseous $^{129}$Xe, and a cross-relaxation and detection system (the left part
of the valve K4). The two parts are connected by a cylindrical Pyrex tube and
separated by stopcocks. The whole system was evacuated with K2, K5 valves
closed and K1, K3 and K4 valves opened. When the vacuum reached to 1.5$\times
$10$^{-5}$ Torr for several hours, the valves K1, K4 and K5 were closed and K2
and K3 opened. The cylindrical pump cell containing a few drops of metal Cs,
with a volume of 600 cm$^{3}$, was loaded 760 Torr natural isotopic xenon gas
at room temperature and an atmosphere, then all valves were closed. The pump
cell, which was placed in a 25 Guass magnetic field generated by Helmholtz
coils, was maintained at approximately 333$\pm$4 K by a resistance heater (not
shown) during the optical pumping. The inner surfaces of the cylindrical Pyrex
tube and the pump cell were coated with silane in order to slow down the
relaxation of the $^{129}$Xe upon collision with the tube wall. As soon as all
valves were closed, laser light from a 15 W cw tunable semiconductor-diode
laser array (Opto Power Co. Model OPC-D015-850-FCPS) at wavelength $\lambda
$=852.1 nm was introduced to the system. After passing through beam expander,
Glan prism, $\lambda$/4 plate and convex lens, the laser light became
circularly polarized, and illuminated almost 4/5 of the pump cell volume. The
propagation of the laser paralleled to the orientation of the magnetic field
B. The circularly polarized laser resonates with the Cs D$_{2}$ absorption
line and induces an electron spin polarization in the Cs atoms via a standard
optical pumping process \cite{Zeng,Happer}. After about 25 minutes, the
hyperpolarized $^{129}$Xe gas was produced by spin-exchange collision with the
Cs atoms.

At the part of cross-relaxation and detection system, there were a liter of
$^{1}$HCl gas at room temperature and an atmosphere in the tank. K4 and K5
valves were opened to allow the hyperpolarized $^{129}$Xe and $^{1}$HCl to be
mixed, then transferred to the NMR sample probe (10 mm diameter, 8 mm inner
diameter), pre-cooled to 172$\pm$2 K, of the Bruker SY-80M NMR spectrometer.
The temperature of the probe was subsequently reduced to 142 K in an evacuated
glass dewar by flowing cold nitrogen gas from liquid nitrogen tank, which
controlled by the Bruker variable temperature unit. This temperature was kept,
and the variation was less than 2 K at all times so that the mixture could be
in the solid-state. Since there exist imbalance of polarization, the
hyperpolarized $^{129}$Xe and the thermally polarized proton, in the mixed
system, the interaction between the $^{129}$Xe and the proton can make spin
exchange occur via SPINOE.

The time dependence of the $^{129}$Xe NMR signal intensity (solid circle)
observed after blending hyperpolarized $^{129}$Xe and $^{1}$HCl is shown in
Figure 2(a). The observed spin-lattice relaxation time of $^{129}$Xe in the
presence of $^{1}$HCl is 29.6$\pm$0.6 min. The initial rise of signal
manifests the accumulation of solid-state hyperpolarized $^{129}$Xe in the
probe, similar to that in Ref. 19. At the peak signal, the increase of
accumulated $^{129}$Xe magnetization reaches a balance with the decay of
$^{129}$Xe magnetization via SPINOE with $^{1}$HCl. After this peak point,
$^{129}$Xe magnetization decays as a result of SPINOE. So the solid line, a
fit to data after the peak, represents the time dependence of $^{129}$Xe
residual magnetization after a SPINOE experiment.

Figure 2(b) displays the time dependence of the solid-state proton NMR signal
(solid circle) from the same run. Both the accumulation of the condensed state
mixture and the cross-relaxation between the proton and the hyperpolarized
$^{129}$Xe contribute to the initial rise of the proton signal. The proton NMR
signal reaches its peak vaule at a time t=280 s after the mixing, and decays
towards its thermal equilibrium value (dashed line) at the same rate as
$^{129}$Xe signal. The solid line represents a fit to data to guide eyes. We
used pulse flip angles of 4$^{0}$ and 9$^{0}$ for $^{129}$Xe and $^{1}$H
respectively, and $^{129}$Xe was performed with a home-bulit probe.

Typcial spectrum for the enhanced proton is shown in Figure 3. In order to
obtain the large $^{1}$H NMR signal, we used 90$^{0}$ pulse angle for a single
acquisition at a time t=280 s. Figure 3(b) presents a typical enhanced $^{1}$H
NMR spectrum comparing with the one at thermal equilibrium (Figure 3(a)). Due
to the relation of the magnetization versus the nuclear spin polarization
given by Abragam \cite{Abragam}, the enhancement factor of the solid-state
$^{1}$H was about 6, which corresponds to the proton polarization of
8.55$\times$10$^{-5},$ on the basis of compariosn of the integrated intensity
of the laser-enhanced signal with that at equilibrium.

\section{Theoretical Analysis and Discussion}

The interactions are complicated in solid-state, but in this paper, we will
limit our discussion to only spins-$\frac{1}{2}$ nuclei coupled via the direct
nuclear dipolar interaction:%

\begin{equation}
H_{D}=\frac{\gamma_{_{S}}\gamma_{_{I}}}{r^{3}}[S\bullet I-3(r\bullet
S)(r\bullet I)],
\end{equation}
where $I$ and $S$ are the spin angular momentum operators, $\gamma_{I}$\ and
$\gamma_{S}$ are the gyromagnetic ratios of I and S spins, respectively, and
$r$ is the distance between two spins. Due to the perturbation of Hamiltonian
$H_{D},$ the transition probabilities $W$ between the eigenstates\ can be
given \cite{Solomon}:%
\begin{equation}%
\begin{array}
[c]{c}%
W_{0}^{IS}=\frac{2\delta}{20}J(\omega_{I}-\omega_{S}),\\
W_{1I}^{IS}=\frac{3\delta}{20}J(\omega_{I}),\\
W_{1S}^{IS}=\frac{3\delta}{20}J(\omega_{S}),\\
W_{2}^{IS}=\frac{12\delta}{20}J(\omega_{I}+\omega_{S}),
\end{array}
\end{equation}
with%
\begin{equation}
\delta=\frac{\hbar^{2}\gamma_{S}^{2}\gamma_{I}^{2}}{r^{6}},
\end{equation}%
\begin{equation}
J(\omega)=\frac{\tau_{c}}{1+\omega^{2}\tau_{c}^{2}},
\end{equation}
here, $\hbar$\ is the Planck constant divided by 2$\pi$, and $\tau_{c}$ is the
correlation time of spin systems.

Accroding to the Solomon equations, the evolution of this two-spin systems,
$^{1}$H (I=$\frac{1}{2}$) and $^{129}$Xe (S=$\frac{1}{2}$), could be described
by \cite{Solomon,Abragam}:%
\begin{equation}
\frac{d}{dt}\left(
\begin{array}
[c]{c}%
I_{z}\\
S_{z}%
\end{array}
\right)  =-\left[
\begin{array}
[c]{cc}%
\rho_{I} & \sigma_{IS}\\
\sigma_{SI} & \rho_{S}%
\end{array}
\right]  \left(
\begin{array}
[c]{c}%
I_{z}-I_{0}\\
S_{z}-S_{0}%
\end{array}
\right)  ,
\end{equation}
where I$_{z}$ and S$_{z}$ are z components of the spins I and S respectively,
I$_{0}$ and S$_{0}$ are their equilibrium values, $\rho_{I}$\ and $\rho_{S}$
are the autorelaxation rates of the $^{1}$H and the $^{129}$Xe spins, and
$\sigma_{IS}$ and $\sigma_{SI}$\ are the corresponding cross-relaxation rates.
The elements of cross-relaxation matrix can be expressed by the transition
probabilities W resulting from II, SS and IS interactions \cite{Ernst}:%

\begin{equation}%
\begin{array}
[c]{c}%
\rho_{I}=2(n_{I}-1)(W_{1}^{II}+W_{2}^{II})+n_{S}(W_{0}^{IS}+2W_{1I}^{IS}%
+W_{2}^{IS}),\\
\rho_{S}=2(n_{S}-1)(W_{1}^{SS}+W_{2}^{SS})+n_{I}(W_{0}^{IS}+2W_{1S}^{IS}%
+W_{2}^{IS}),\\
\sigma_{IS}=n_{S}(W_{2}^{IS}-W_{0}^{IS}),\\
\sigma_{SI}=n_{I}(W_{2}^{IS}-W_{0}^{IS}),
\end{array}
\end{equation}
here, $n_{I}$ and $n_{s}$\ are magnetically equivalent I and S spins. A
solution of equation (5) of particular interest in this system is the one
corresponding to the initial conditions:%
\begin{equation}%
\begin{array}
[c]{c}%
(I_{z}-I_{0})_{t=0}=0,\\
(S_{z}-S_{0})_{t=0}=S_{i}.
\end{array}
\end{equation}
So the solution can be given by:%
\begin{equation}%
\begin{array}
[c]{c}%
I_{z}(t)=I_{0}+C(\exp(\lambda_{1}\cdot t)-\exp(\lambda_{2}\cdot t)),\\
S_{z}(t)=S_{0}+C[r_{1}\cdot(\exp(\lambda_{1}\cdot t)-r_{2}\cdot\exp
(\lambda_{2}\cdot t))],
\end{array}
\end{equation}
where $\lambda_{1}$\ and $\lambda_{2}$ are given by:%
\begin{equation}%
\begin{array}
[c]{c}%
\lambda_{1}=\frac{-(\rho_{I}+\rho_{S})-\sqrt{(\rho_{I}-\rho_{S})^{2}%
+4\sigma_{IS}\sigma_{SI}}}{2},\\
\lambda_{2}=\frac{-(\rho_{I}+\rho_{S})+\sqrt{(\rho_{I}-\rho_{S})^{2}%
+4\sigma_{IS}\sigma_{SI}}}{2},
\end{array}
\end{equation}
and%
\begin{equation}%
\begin{array}
[c]{c}%
r_{1}=\frac{\rho_{S}-\sqrt{(\rho_{I}-\rho_{S})^{2}+4\sigma_{IS}\sigma_{SI}}%
}{2\sigma_{IS}},\\
r_{2}=\frac{\rho_{S}+\sqrt{(\rho_{I}-\rho_{S})^{2}+4\sigma_{IS}\sigma_{SI}}%
}{2\sigma_{IS}},
\end{array}
\end{equation}%
\begin{equation}
C=\frac{S_{i}}{r_{1}-r_{2}}.
\end{equation}
Therefore, the $^{1}$H enhancemet via cross-relaxation comparing with that at
thermal equilibrium is:%
\begin{equation}
\frac{I_{z}(t)-I_{0}}{I_{0}}=-\frac{\sigma_{IS}}{\rho_{I}}\frac{\gamma_{S}%
}{\gamma_{I}}\frac{S_{z}(t)-S_{0}}{S_{0}},
\end{equation}
where $[S_{z}(t)-S_{0}]/S_{0}$, the enhancement of hyperpolarized solid-state
$^{129}$Xe, is about 6000 in our experiment, which corresponds to the $^{129}%
$Xe polarization of 2.16\% \cite{Zhou}. Inserting the values (2) and (6) into
(12) we get:%

\begin{equation}
\frac{I_{z}(t)-I_{0}}{I_{0}}=\frac{-\gamma_{S}(S_{z}(t)-S_{0})n_{S}[6\delta
J(\omega_{I}+\omega_{S})-\delta J(\omega_{I}-\omega_{S})]}{\gamma_{I}%
S_{0}\{(n_{I}-1)[3\delta^{\prime}J(\omega_{I})+12\delta^{\prime}J(2\omega
_{I})]+n_{S}[\delta J(\omega_{I}-\omega_{S})+3\delta J(\omega_{I})+6\delta
J(\omega_{I}+\omega_{S})]\}},
\end{equation}
with%
\begin{equation}
\delta^{\prime}=\frac{\hbar^{2}\gamma_{I}^{4}}{r^{\prime6}},
\end{equation}
here $r^{\prime}$ is the distance between two neighboring I spins.

In our experiment, $\omega_{I}$ ($^{1}$H) and $\omega_{S}$\ ($^{129}$Xe)\ are
80.13MHz and 22.16MHz (on Bruker AC-80 spectrometer). The concentrations of
$^{1}$H and $^{129}$Xe are 34.6mmol/cm$^{3}$ and 5.46mmol/cm$^{3}$
respectively. We can image a model that one $^{129}$Xe nucleus is surrounded
by six $^{1}$H nuclei on average. The autorelaxation rate of the proton
$\rho_{I}$\ of (66 s)$^{-1}$ is measured in the experiment so that one can
obtain the cross relaxation $\sigma_{IS}$\ by the equation (12). Therefore,
the correlation time $\tau_{c}$\ is estimated to be 6.18$\times$10$^{-5}$\ s.
Using the equation (6), we can calculate that autorelaxation rates $\rho_{I}%
$\ and $\rho_{S}$\ are (81 s)$^{-1}$ and (62 min)$^{-1}$, and cross relaxation
rates $\sigma_{IS}$\ and $\sigma_{SI}$\ are 5.53$\times$10$^{-5}$ s$^{-1}$ and
3.32$\times$10$^{-4}$ s$^{-1}$, respectively. Thus from equation (8), the time
evolution of the proton polarization can be theoretically writen as:%
\begin{equation}
I_{z}(t)=I_{0}+C(\exp(-t/3744)-\exp(-t/81)).
\end{equation}
The theoretical simulation (dot line) is visualized in Figure 2(b) for clear
comparison to experimental results. Although only the direct nuclear dipolar
interaction is considered in the above discussion, in fact, there are so many
factors, such as spin rotation, paramagnetisim, relaxation with the wall,
inhomogenuous magnetic field, etc., which can induce the relaxation.
Consequently the theoretical relaxation rates are smaller than the
experimental ones, and predictions of the time dependence of the proton
polariztion are larger than experimental results.

Assuming $r$\ is equal to $r^{\prime}$\ for simplification in our theoretical
model, one can calculate the maximum solid-state proton enhancement of 7.1
using equation (13), which is in general agreement with the measured value.
Although it is substantially smaller than the $^{129}$Xe\ enhancement, this
can not indicate the possibility that only a fraction of the total number of
$^{1}$HCl molecules interact with the hyperpolarized $^{129}$Xe by
diploar-diploar. Because only the natural xenon (26.4\% enriched $^{129}$Xe,
21.2\% enriched $^{131}$Xe) has been used, at the temperature of 142K, the
cross-relaxation between $^{129}$Xe and the isotope $^{131}$Xe can decrease
the efficiency of polarization transfer from $^{129}$Xe to $^{1}$H
\cite{Gatzke}. On the other hand, dipolar relaxation from diffusing vacancies
dominates at this temperature for $^{129}$Xe spin-lattice relaxation
\cite{Cates}. If we take into account of all above factors, together with the
inhomogenuous magnetic field due to the fluctuation of eletromagnet and the
polarization loss during the phase transition, etc., all of these losses maybe
counterbalance the difference between experimental values and theoretical ones.

According to equation (13), for further increasing the proton enhancement, one
should first obtain the greatest $^{129}$Xe enhancement via employing the high
power and narrow bandwidth laser, and/or increase the gas pressure of the pump
cell in order to enhance the optical-pumped absorbed power when using the wide
bandwidth laser. A small proton number density by using partially deuterated
sample and a large $^{129}$Xe number density by using Xe isotopically enriched
with $^{129}$Xe also can boost transfer efficiency.

\section{Conclusion}

\ \ In conclusion, we firstly obtained the solid-state laser-enhanced NMR
proton signal of $^{1}$HCl via SPINOE by cooling hyperpolarzied $^{129}$Xe to
solid-state at 1.87T and 142K in the flow system. The amplification factor of
the proton nuclear polarization is about 6. The theoretical calculated
enhancement is in agreement with the measured value. Furthermore, the
hyperpolarized $^{129}$Xe has long lifetime in the solid-state, which means
that $^{1}$H could keep the longer coherence time with hyperpolarized $^{129}%
$Xe. Therefore this method may be useful to overcome one of the difficult
problems in liquid-state NMR quantum computer and to establish solid-state
quantum computers.

Solid-state signal enhancement via SPINOE with hyperpolarized $^{129}$Xe is
not limited to the proton, i.e., it can be expanded to other nuclei. Although
we have focused on quantum computers, this method should be readily extended
to material science and determination of the 3D structure of large
biomolecules, since NOE can provide unique information on molcular structure
which can not be obtained with any other known technique. Traditional NOE
needs to irradiate one nucleus in order to observe anther. However, it is very
convenient and available for probing interactions between nuclei without any
additional conditions by using this method.

\section{Acknowledgment}

\ \ \ This work is supported by the National Natural Science Foundation of
China under Grant No. 10234070, National Science Fund for Distinguished Young
Scholars under Grant No. 29915515 and National Fundamental Research Program
under Grant No. 2001CB309306.

\begin{center}
{\LARGE Figure captions}
\end{center}

\bigskip

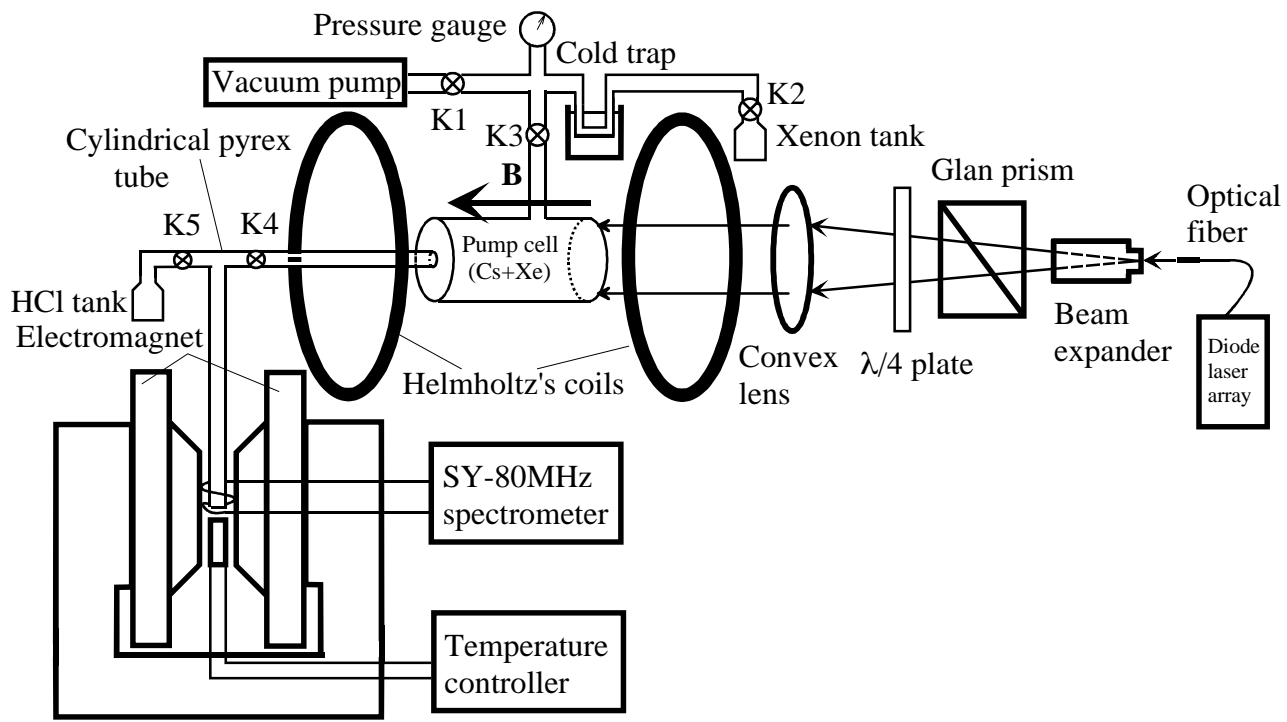
Fig. 1. Schematic diagram of the experimental setup.

Fig. 2. (a) Time dependence of the solid-state $^{129}$Xe NMR signal (solid
circle) observed after the blend of hyperpolarized $^{129}$Xe and $^{1}$HCl at
142 K and 1.87 T. The solid line, a fit to data after the peak, represents how
much $^{129}$Xe magnetization is left after a SPINOE experiment. (b) Time
dependence of the solid-state proton NMR signal (solid circle) from the same
sample, and it relaxes towards its thermal equilibrium value (dashed line).
The dot line represents the theoretical predictions for comparison to
experimental results (solid line).

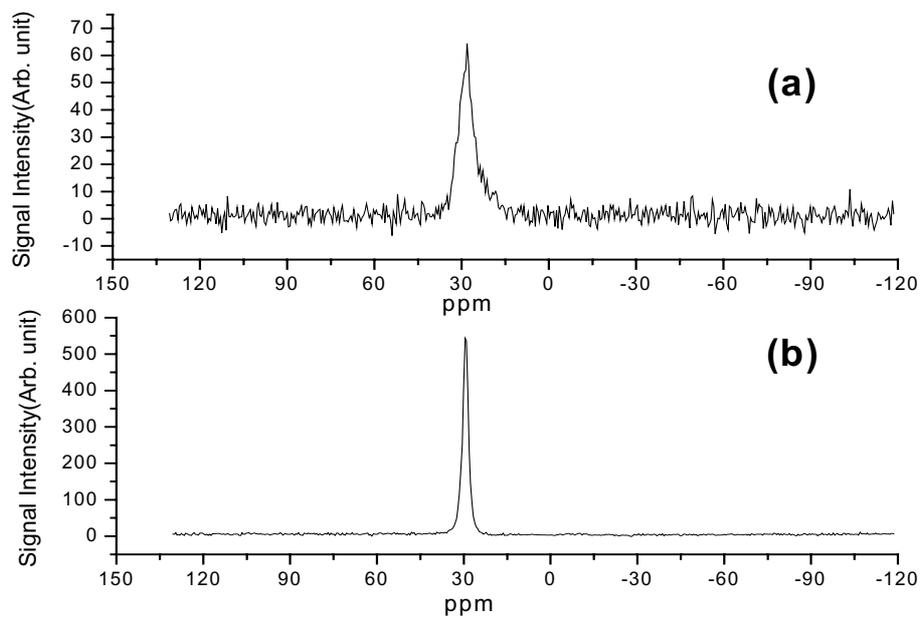
Fig. 3. (a) The solid-state proton NMR spectra of $^{1}$HCl under the
conditions of thermal equilibrium and (b) laser-enhanced signal via SPINOE
with hyperpolarized $^{129}$Xe.

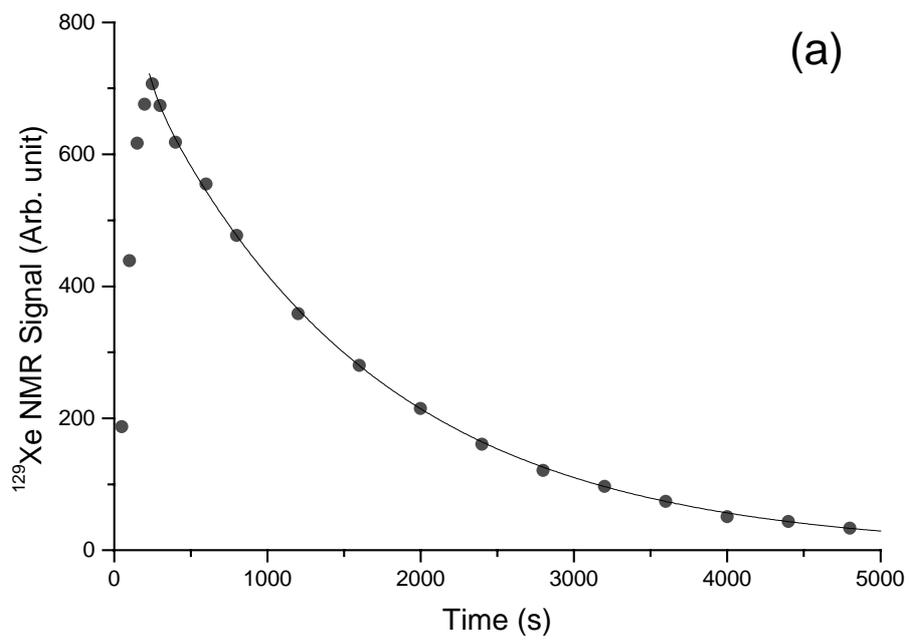
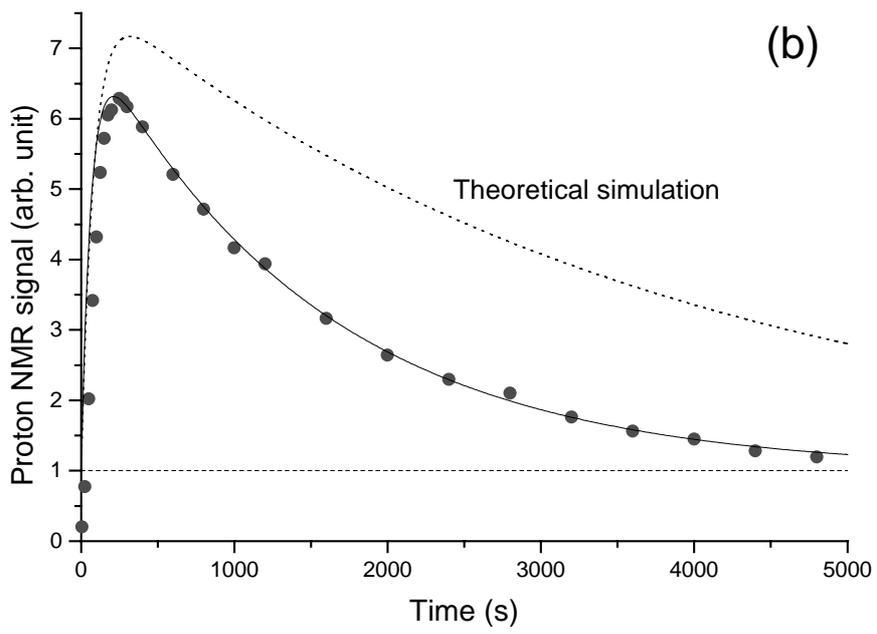
\end{document}